\documentclass{elsart}

\usepackage{graphicx}

\begin{document}
\begin{frontmatter}

\title{Triangular arbitrage as an interaction \\ among foreign exchange rates}

\author{Yukihiro Aiba$^{\rm a,1}$, Naomichi Hatano$^{\rm a}$, Hideki Takayasu$^{\rm b}$,}
\author{Kouhei Marumo$^{\rm c}$, Tokiko Shimizu$^{\rm d}$}

\thanks{e-mail: aiba@phys.aoyama.ac.jp}
\address{$^{\rm a}$Department of Physics, Aoyama Gakuin University\\
Chitosedai 6-16-1, Setagaya, Tokyo 157-8572, Japan}
\address{$^{\rm b}$Sony CSL, Higashi-Gotanda 3-14-13,\\
Shinagawa, Tokyo 141-0022, Japan}
\address{$^{\rm c}$Institute for Monetary and Economic Studies, Bank of Japan,\\
Hongoku-cho Nihonbashi 2-1-1, Chuo, Tokyo 103-8660, Japan}
\address{$^{\rm d}$Financial Markets Department, Bank of Japan,\\
Hongoku-cho Nihonbashi 2-1-1, Chuo, Tokyo 103-8660, Japan}


\begin{abstract}
We first show that there are in fact triangular arbitrage opportunities in the spot foreign exchange markets, analyzing the time dependence of the yen-dollar rate, the dollar-euro rate and the yen-euro rate.
Next, we propose a model of foreign exchange rates with an interaction. 
The model includes effects of triangular arbitrage transactions as an interaction among three rates. 
The model explains the actual data of the multiple foreign exchange rates well. 
\end{abstract}

\end{frontmatter}

\section{Introduction}

The triangular arbitrage is a financial activity that takes advantage of the three exchange rates among three currencies.
Suppose that we exchange one US dollar to some amount of Japanese yen, exchange the amount of Japanese yen to some amount of euro, and finally exchange the amount of euro back to US dollar; then how much US dollar do we have? There are opportunities that we have more than one US dollar. The triangular arbitrage transaction is the trade that takes this type of opportunities. It has been argued that the triangular arbitrage makes the product of the three exchange rates converge to a certain value \cite{moosa}. In other words, the triangular arbitrage is a form of interaction among currencies. 

The purpose of this paper is to show that there is in fact triangular arbitrage opportunities in foreign exchange markets and they generate an interaction among foreign exchange rates. 
We first analyze real data in Sec.\ \ref{sec:observation}, showing that the product of three foreign exchange rates has a narrow distribution with fat tails. 
In order to explain the behavior, we propose in Sec.\ \ref{sec:modeling} a model of the time evolution of exchange rates with an interaction.
The simulation explain the real data well.

\section{Observation of the Triangular Arbitrage Opportunity in High-Frequency Data}
\label{sec:observation}

In the present paper, we analyze actual data of the yen-dollar rate, the yen-euro rate and the dollar-euro rate, taken from January 25 1999 to March 12 1999 except for weekends.
We show in this section that there are actually triangular arbitrage opportunities and that the three exchange rates correlate strongly.

\subsection{Existence of Triangular Arbitrage Opportunities} 

In order to quantify the triangular arbitrage opportunities, we define the quantity
\begin{equation}
  \label{eq:prod}
  \mu(t) = \prod_{i=1}^{3}r_i(t)\ ,
\end{equation} 
where $r_i(t)$ denotes each exchange rate at time $t$. We refer to this quantity as the rate product. There is a triangular arbitrage opportunity whenever the rate product is greater than unity.

To be more precise, there are two types of the rate product. One is based on the arbitrage transaction in the direction of dollar to yen to euro to dollar. The other is based on the transaction in the opposite direction of dollar to euro to yen to dollar. Since these two values show similar behavior, we focus on the first type of $\mu(t)$ hereafter. Thus, we specifically define each exchange rate as
\begin{eqnarray}
r_1(t)&\equiv& \frac{1}{\mbox{yen-dollar ask }(t)}\\
r_2(t)&\equiv& \frac{1}{\mbox{dollar-euro ask }(t)}\\
r_3(t)&\equiv& \mbox{yen-euro bid }(t).
\end{eqnarray}
(Note the difference between
\begin{equation}
\begin{array}{ccccc}
1&[dollar]&\rightarrow&\mbox{yen-dollar bid}                 &[yen]
\end{array}
\end{equation}
and
\begin{equation}
\begin{array}{ccccc}
1&[yen]   &\rightarrow&{\displaystyle\frac{1}{\mbox{yen-dollar ask}}}&[dollar].{\mbox{)}}
\end{array}
\end{equation}
We assume here that an arbitrager can transact instantly at the bid and the ask prices provided by information companies and hence we use the prices at the same time to calculate the rate product.

\begin{figure}[p]
(a)
    \begin{center} 
      \includegraphics[width=\textwidth]{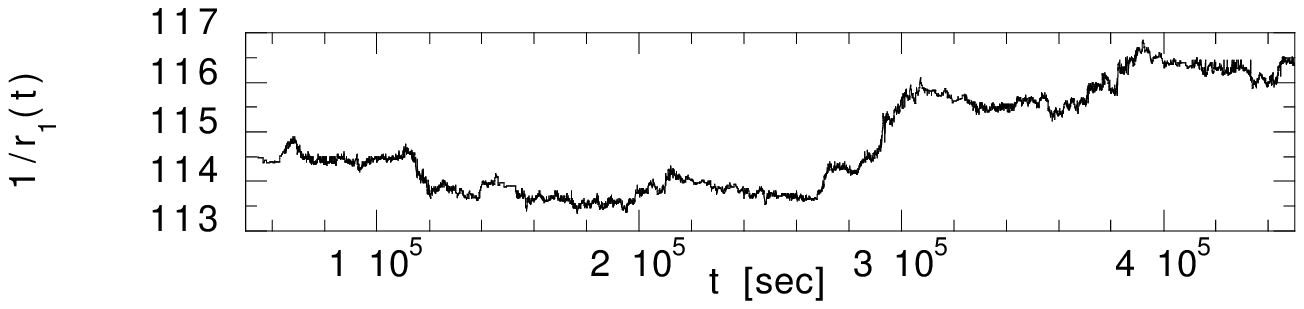} 
    \end{center}
(b)
    \begin{center} 
      \includegraphics[width=\textwidth]{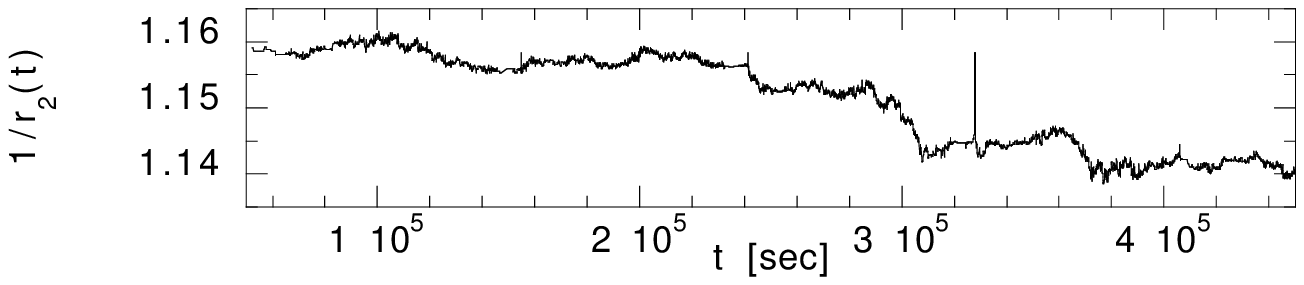} 
    \end{center}
(c)
    \begin{center} 
      \includegraphics[width=\textwidth]{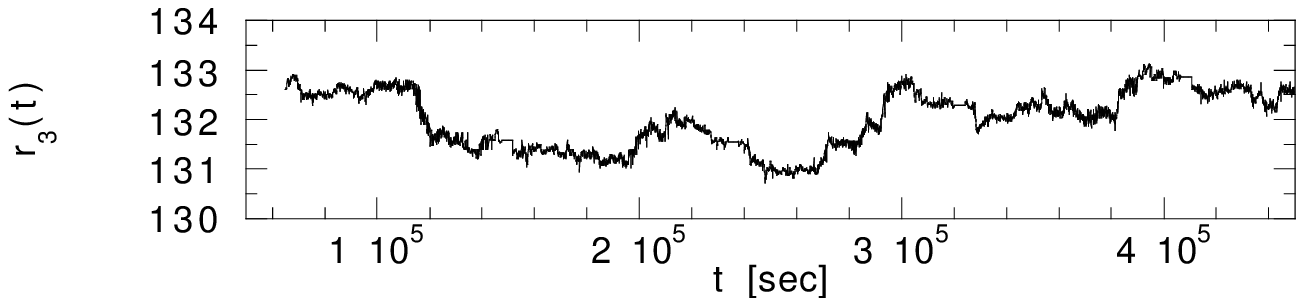} 
    \end{center}
(d)
    \begin{center} 
      \includegraphics[width=\textwidth]{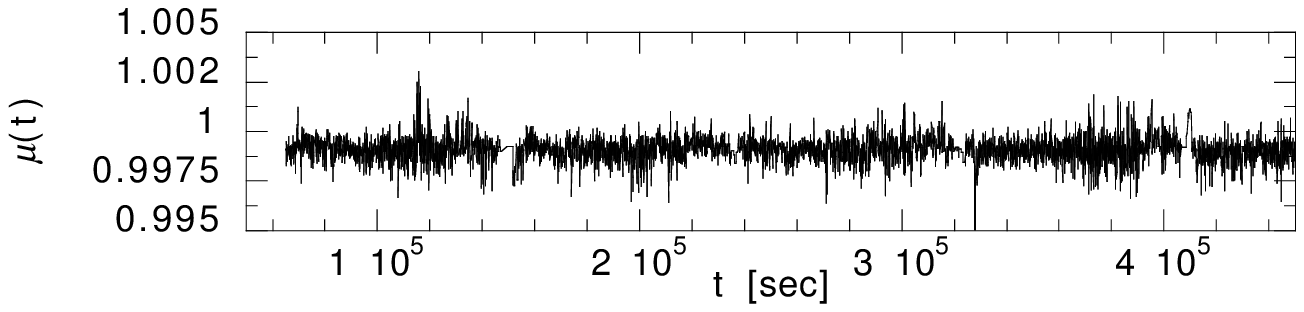} 
    \end{center}
  \caption{The time dependence of  
  (a) the yen-dollar ask $1/r_1$, (b) the dollar-euro ask $1/r_2$, (c)  the yen-euro bid $r_3$ and (d) the rate product $\mu$. The horizontal axis denotes the seconds from 00:00:00, January 25 1999.}
  \label{fig:bidask}
\end{figure}

\begin{figure}[p]
(a) 
 \begin{center}
   \includegraphics[width=8cm]{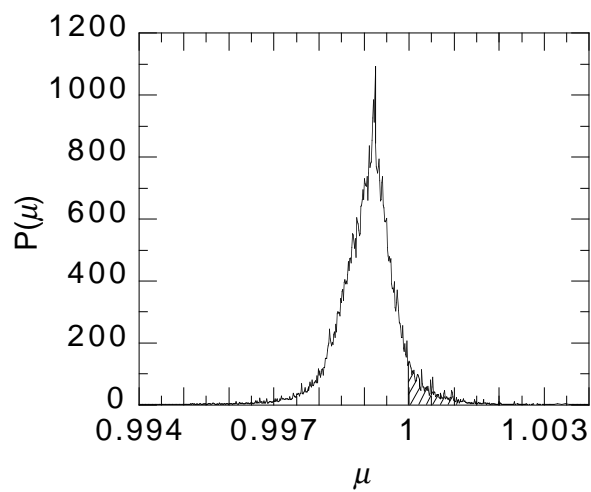} \\
 \end{center}

(b)
  \begin{center}
  \includegraphics[width=8cm]{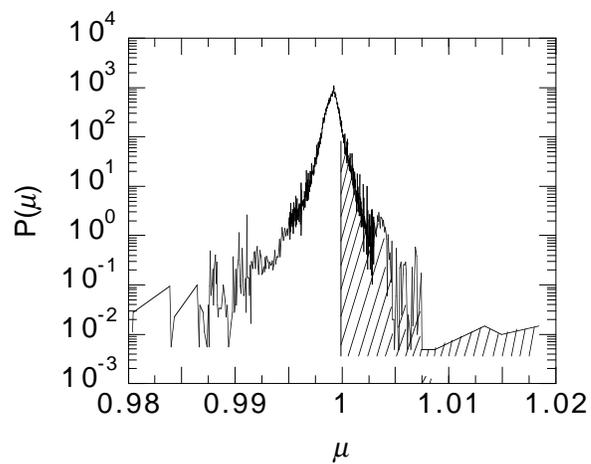} 
  \end{center}
  \caption{The probability density function of the rate product $\mu$. (b) is a semi-logarithmic plot of (a). The shaded area represents triangular arbitrage opportunities. The data were taken from January 25 1999 to  March 12 1999.}
  \label{fig:fils}
\end{figure}

\begin{figure}[p]
(c)
 \begin{center}
   \includegraphics[width=8cm]{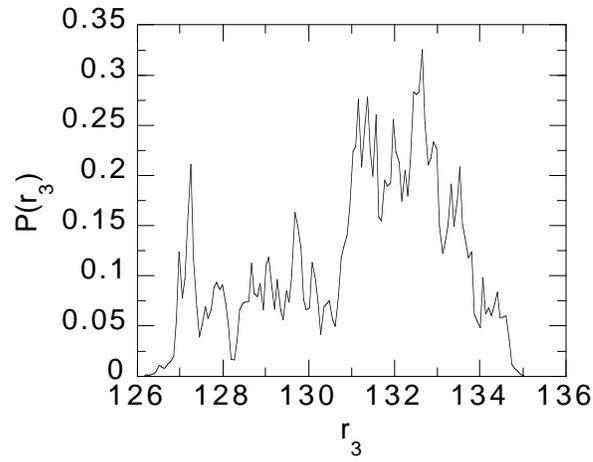} 
 \end{center}
  \caption{The probability density function of the three rates: (a) the reciprocal of the yen-dollar ask, $r_1$, (b) the reciprocal of the dollar-euro ask, $r_2$ and (c) the yen-euro bid $r_3$. The data were taken from January 25 1999 to March 12 1999.}
  \label{fig:ratefils}
\end{figure}

Figure\ \ref{fig:bidask}(a)-(c) shows the actual changes of the three rates: the yen-euro ask, the dollar-euro ask and the yen-euro bid. 
Figure\ \ref{fig:bidask}(d) shows the behavior of the rate product $\mu(t)$.
We can see that the rate product $\mu$ fluctuates around the average 
\begin{equation}
  \label{eq:defep}
  m \equiv \langle\mu(t)\rangle \simeq 0.99998.
\end{equation}
(The average is less than unity because of the spread; the spread is the difference between the ask and the bid prices and is usually of the order of $0.05\%$ of the prices.)
The probability density function of the rate product $\mu$ (Fig.\ \ref{fig:fils}) has a sharp peak and  fat tails while those of the three rates (Fig.\ \ref{fig:ratefils}) do not.
It means that the fluctuations of the exchange rates have correlation that makes the rate product converge to the average $m$.

\subsection{Feasibility of the Triangular Arbitrage Transaction}

We discuss here the feasibility of the triangular arbitrage transaction. We analyze the duration of the triangular arbitrage opportunities and calculate whether an arbitrager can make profit or not. 

The shaded area in Fig.\ \ref{fig:fils} represents triangular arbitrage opportunities. 
We can see that the rate product is grater than unity for about $6.4\%$ of the time.
It means that triangular arbitrage opportunities exist about ninety minutes a day. The ninety minutes, however, include the cases where the rate product $\mu$ is greater than unity very briefly.
The triangular arbitrage transaction is not feasible in these cases. 

\begin{figure}[p]
 \begin{center}
  \includegraphics[width=8cm]{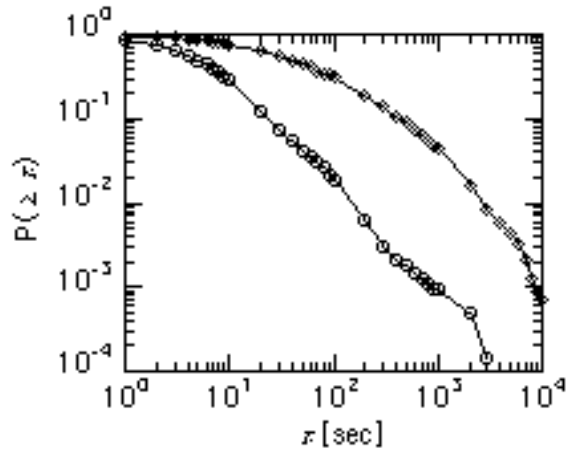} 
 \end{center}
 \caption{The cumulative distributions of $\tau_+$ ($\circ$) and $\tau_-$ ($\diamond$). The distribution of $\tau_+$ shows a power-low behavior. The data were taken from January 25 1999 to March 12 1999.}
 \label{fig:dur}
\end{figure}

In order to quantify the feasibility, we analyze the duration of the triangular arbitrage opportunities.
Figure\ \ref{fig:dur} shows the cumulative distributions of the duration $\tau_{+}$ of the situation $\mu>1$ and $\tau_{-}$ of $\mu<1$. 
It is interesting that the distribution of $\tau_{+}$ shows a power-law behavior while the distribution of $\tau_{-}$ dose not. This difference may suggest that the triangular arbitrage transaction is carried out indeed.

\begin{figure}[p]
 \begin{center}
  \includegraphics[width=8cm]{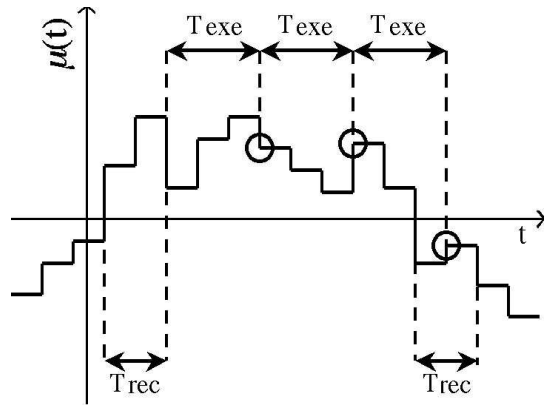} 
 \end{center}
 \caption{A conceptual figure of the profit calculation. We assume that it takes an arbitrager $T_{\rm rec}$[sec] to recognize triangular arbitrage opportunities and $T_{\rm exe}$[sec] to execute a triangular arbitrage transaction. The circles ($\circ$) indicate the instances where triangular arbitrage transactions are carried out.}
 \label{fig:concept}
\end{figure}

\begin{figure}[p]
 \begin{center}
  \includegraphics[width=8cm]{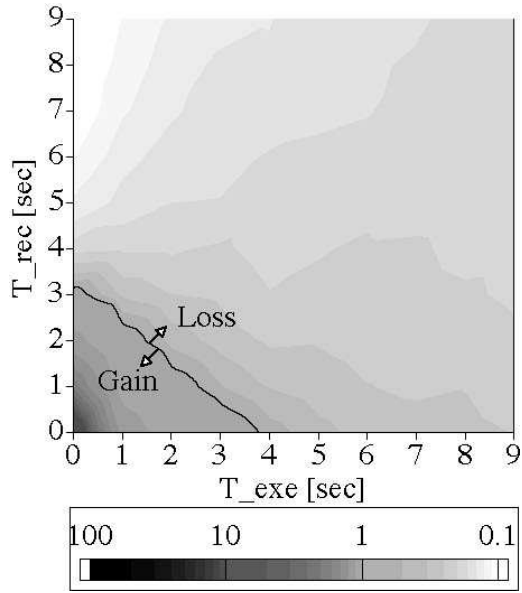} 
 \end{center}
 \caption{ A phase diagram of the profit that an arbitrager can make from one US dollar (or Japanese yen or euro) in a day under the assumption shown in Fig.\ \ref{fig:concept}.}
 \label{fig:gain}
\end{figure}

In order to confirm the feasibility of the triangular arbitrage, we simulate the triangular arbitrage transaction using our time series data.
We assume that it takes an arbitrager $T_{\rm rec}$[sec] to recognize triangular arbitrage opportunities and $T_{\rm exe}$[sec] to execute a triangular arbitrage transaction; see Fig.\ \ref{fig:concept}.
We also assume that the arbitrager transacts whenever the arbitrager recognizes the opportunities.
Figure\ \ref{fig:gain} shows how much profit the arbitrager can make from one US dollar (or Japanese yen or euro) in a day. We can see that the arbitrager can make profit if it takes the arbitrager a few seconds to recognize the triangular arbitrage opportunities and to execute the triangular arbitrage transaction.  

\section{Modeling}
\label{sec:modeling}

We here introduce a new model that takes account of the effect of the triangular arbitrage transaction as an interaction among the three rates.
Many models of price change have been introduced so far:
for example, the L$\acute{{\rm e}}$vy-stable non-Gaussian model \cite{mandelbrot}; the truncated L$\acute{{\rm e}}$vy flight \cite{econophysics}; the ARCH/GARCH processes \cite{arch,garch}. 
They discuss, however, only the change of one price.
They did not consider an interaction among multiple prices. As we discussed in Sec.\ \ref{sec:observation}, however, the triangular arbitrage opportunity exists in the market and is presumed to affect price fluctuations in the way the rate product tends to converge to a certain value. 

\subsection{Basic Time Evolution}\label{subsec:RC}

The basic equation of our model is a time-evolution equation of the logarithm of each rate: 
\begin{equation}
  \label{eq:time.ev.lnr}
  \ln{r_i(t+\Delta t)} = \ln{r_i(t)}+f_i(t)+g(\nu(t)), \ \ \ (i=1,2,3)
\end{equation}
where $\nu$ is the logarithm of the rate product
\begin{equation}
  \label{eq:defnu}
  \nu(t) \equiv \ln\mu(t) = \sum_{i=1}^{3}\ln r_i(t) .
\end{equation}
Just as $\mu$ fluctuates around $m=\langle\mu\rangle\simeq 0.99998$, the logarithm rate product $\nu$ fluctuates around 
\begin{equation}
  \label{eq:epsilon}
  \epsilon\equiv\langle\ln\mu\rangle\simeq -0.00091
\end{equation} 
(Fig.\ \ref{fig:numodel}(a)).
In this model, we focus on the logarithm of the rate-change ratio $\ln(r_i(t+\Delta t)/r_i(t))$, because the relative change is presumably more essential than the absolute change. 
We assumed in Eq.\ (\ref{eq:time.ev.lnr}) that the change of the logarithm of each rate is given by an independent fluctuation $f_i(t)$ and an attractive interaction $g(\nu)$. 
The triangular arbitrage is presumed to make the logarithm rate product $\nu$ converge to the average $\epsilon$; thus, the interaction function $g(\nu)$ should be negative for $\nu$ greater than
$\epsilon$ and positive for $\nu$ less than $\epsilon$: 
\begin{eqnarray}
  \label{eq:negaposi}
  g(\nu) 
  \left\{ 
  \begin{array}{ccc}
  < 0 &,& \mbox{for } \nu > \epsilon \\
  > 0 &,& \mbox{for } \nu < \epsilon  .
  \end{array}         
  \right.
\end{eqnarray}
As a linear approximation, we define $g(\nu)$ as
\begin{equation}
  \label{eq:def.gu}
  g(\nu)\equiv -a(\nu-\epsilon) 
\end{equation}
where $a$ is a positive constant which specifies the interaction strength.

The time-evolution equation of $\nu$ is given by summing Eq.\ (\ref{eq:time.ev.lnr}) over all $i$:
\begin{equation}
\label{eq:nuevo}
\nu(t+\Delta t)-\epsilon =(1- 3a)(\nu(t)-\epsilon)+F(t),
\end{equation}
where
\begin{equation}
  \label{eq:Ft}
F(t)\equiv \sum_{i=1}^3 f_{i}(t).  
\end{equation}
This is our basic time-evolution equation of the logarithm rate product.

\subsection{Estimation of Parameters}

\begin{figure}[p]
(a)
    \begin{center}
      \includegraphics{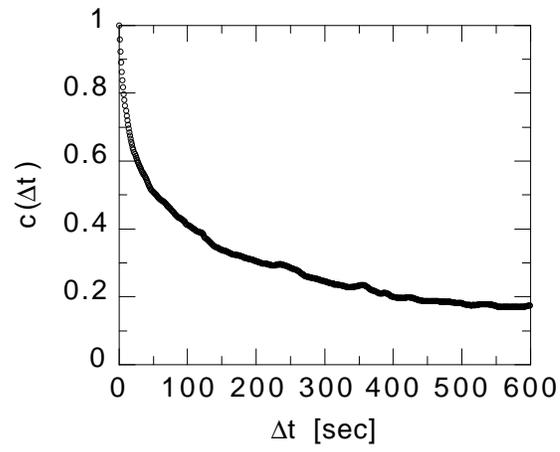} 
    \end{center}
(b)
    \begin{center}
      \includegraphics{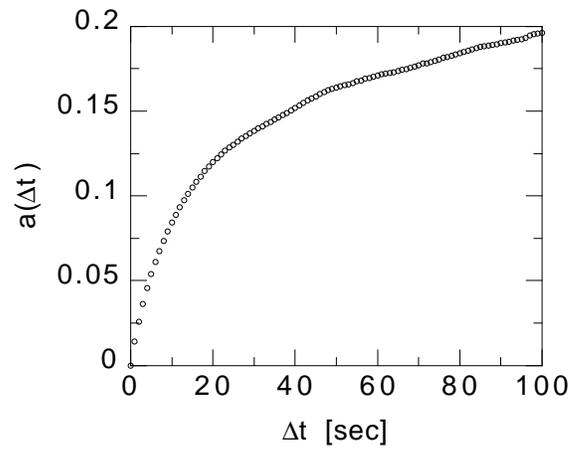} 
    \end{center}
  \caption{(a) The auto-correlation function of $\nu$, $c(\Delta t)$. (b) The time-step dependence of $a(\Delta t)$.} 
  \label{fig:cadt}
\end{figure}

The interaction strength $a$ is related to the auto-correlation function of $\nu$ as follows:
\begin{equation}
  \label{eq:adt}
  1-3a = c(\Delta t) \equiv \frac{ \langle \nu(t+\Delta t) \nu(t) \rangle -
  \langle \nu(t) \rangle ^2}{ \langle \nu ^2 (t) \rangle - \langle
  \nu(t) \rangle ^2} .
\end{equation}
Using Eq.\ (\ref{eq:adt}), we can estimate $a(\Delta t)$ from the real data series as a function of the time step $\Delta t$.  
The auto-correlation function $c(\Delta t)$ is shown in Fig.\ \ref{fig:cadt}(a).  The estimate of $a(\Delta t)$ is shown in Fig.\ \ref{fig:cadt}(b). Hereafter, we fix the time step at $\Delta t = 60$[sec] and hence use 
\begin{equation}
  \label{eq:a1min}
  a(1[\rm{min}])=0.17\pm 0.02
\end{equation}
for our simulation. 

\begin{figure}[p]
(c)
 \begin{center}
   \includegraphics{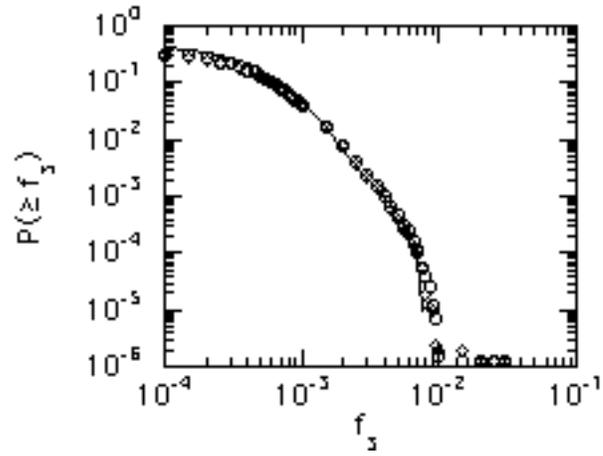} 
 \end{center}
  \caption{The cumulative distributions of the one-minute changes $|\ln r_i(t+1[\mbox{min}])-\ln r_i(t)|$ ($\circ$ represents upward movements and $\diamond$ represents downward movements) and the generated noise $f_i$ (---): (a) the yen-dollar ask and $f_1$, (b) the dollar-euro ask and $f_2$, and (c) yen-euro bid and $f_3$. The real data were taken from January 25 1999 to March 12 1999.}
  \label{fig:delta}
\end{figure}

On the other hand, the fluctuation of foreign exchange rates is known to be a fat-tail noise \cite{bouchaud,takayasu}.
Here we take $f_i(t)$ as the truncated L$\rm\acute{e}$vy process \cite{econophysics,mantegna}:
\begin{eqnarray}
P_{\rm TLF}(f;\alpha,\gamma,l)=
qP_{\rm L}(f;\alpha,\gamma) \Theta(l-|f|),\\
\end{eqnarray}
where $q$ is the normalization constant, $\Theta(x)$ represents the step function and $P_{\rm L}(x;\alpha,\gamma)$ is the symmetric L$\rm\acute{e}$vy distribution of index $\alpha$ and scale factor $\gamma$:
\begin{equation}
P_{\rm L}(x;\alpha,\gamma)=\frac{1}{\pi}\int_{0}^{\infty} e^{-\gamma|k|^\alpha}\cos(k x) {\rm d} k\quad 0<\alpha<2.
\end{equation}
We determine the parameters $\alpha$, $\gamma$ and $l$ by using the following relations for $1<\alpha<2$ \cite{bouchaud,voit}:
\begin{eqnarray}
  \label{eq:c2}
c_2 &=& \frac{\alpha(\alpha -1)\gamma}{|\cos(\pi \alpha/2)|}l^{2-\alpha} , \\
  \label{eq:kappa}
\kappa &=& \frac{(3-\alpha)(2-\alpha)|\cos(\pi \alpha/2)|}{\alpha(\alpha-1)\gamma}l^\alpha ,
\end{eqnarray}
where $c_n$ denotes the $n$th cumulant and $\kappa$ is the kurtosis $\kappa=c_4/{c_2}^2$. The estimates are shown in Table\ \ref{tb:para}.
\begin{table}
  \begin{center}
    \begin{tabular}{l|ccc}
      \hline
      \hline
      rate & $\alpha$ & $\gamma$ & $l$ \\ \hline
      $r_1$ (1/yen-dollar ask)  & 1.8 & 7.61$\times10^{
      -7}$
      & 1.38$\times10^{-2}$ \\
      $r_2$ (1/dollar-euro ask) & 1.7 & 4.06$\times10^{
      -7}$ & 3.81$\times10^{-2}$ \\
      $r_3$ (yen-euro bid)      & 1.8 & 6.97$\times10^{
      -7}$ & 7.58$\times10^{-2}$ \\ \hline\hline
    \end{tabular}
  \end{center}
  \caption{The estimates of the parameters.}
  \label{tb:para}
\end{table}
The generated noises with the estimated parameters are compared to the actual data in Fig.\ \ref{fig:delta}.

\begin{figure}[p]
(a)
  \begin{center}
  \includegraphics{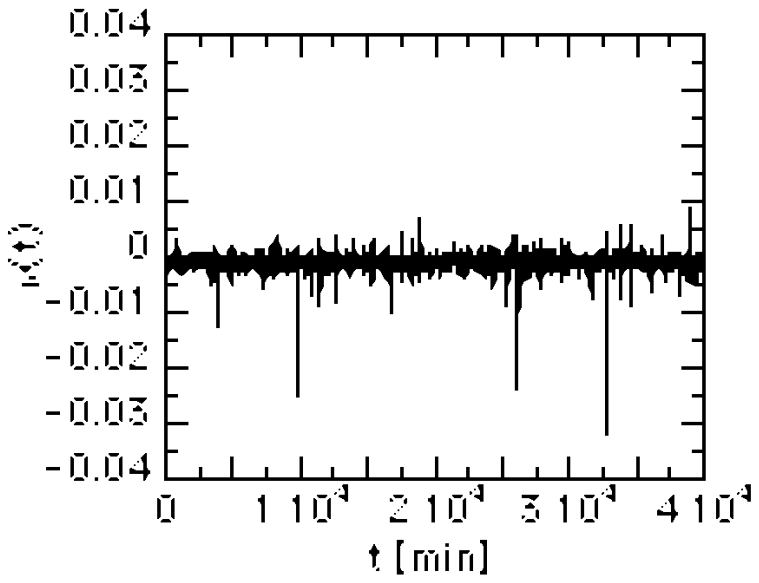} 
  \end{center}
(b)
  \begin{center}
  \includegraphics{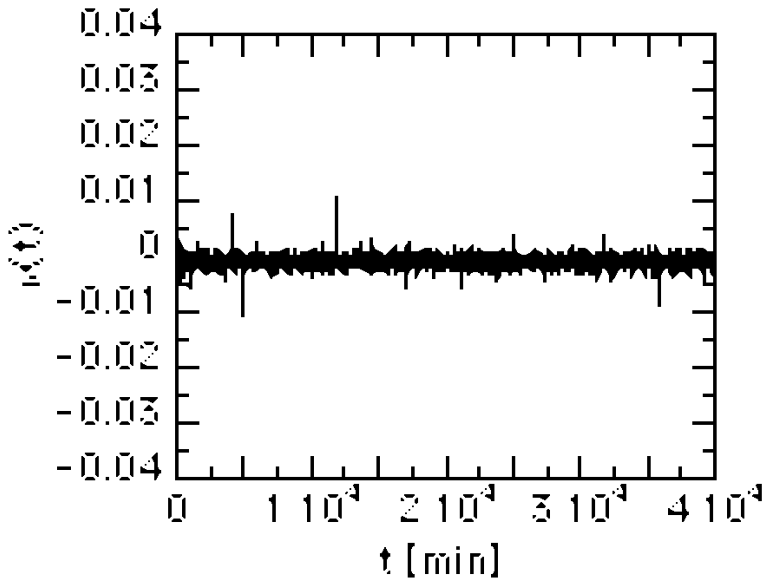} 
  \end{center}
(c)
  \begin{center}
  \includegraphics{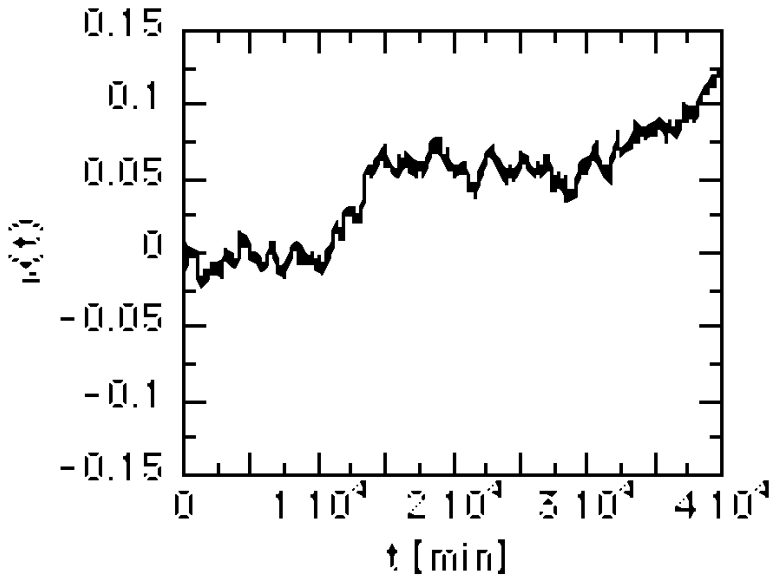} 
  \end{center}
  \caption{The time dependence of $\nu(t[\rm{min}])$ of (a) the real data, (b) the simulation data with the interaction and (c) without the interaction. In (b), $\nu$ fluctuates around $\epsilon$ like the real data.} 
  \label{fig:numodel}
\end{figure}

\begin{figure}[p]
  \begin{center}
  \includegraphics{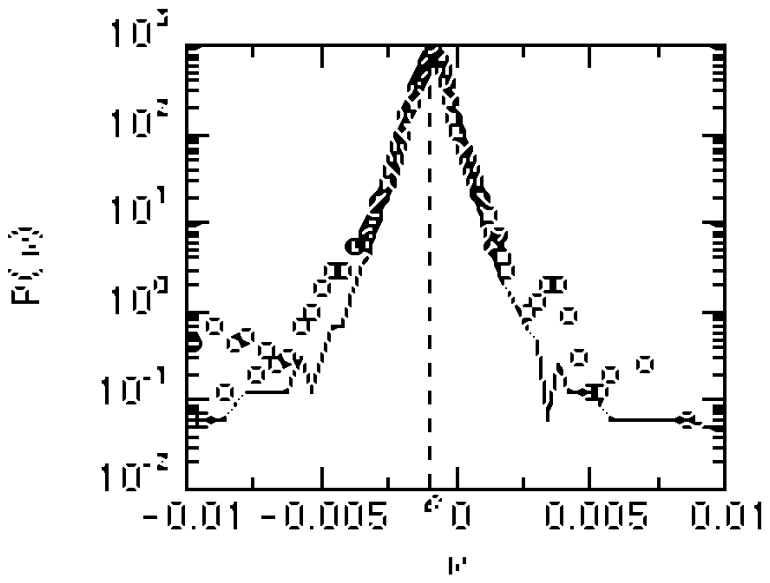} 
  \end{center}
  \caption{The probability density function of $\nu$. 
    The circle ($\circ$) denotes the real data
    and the solid line denotes our simulation data with
     the interaction. The
    simulation data fit the real data well. }
  \label{fig:compfil}
\end{figure}

We simulated the time evolution\ (\ref{eq:nuevo}) with the parameters given in Eqs.\ (\ref{eq:epsilon}),\ (\ref{eq:a1min}) and Table 1.
The probability density function of the results (Fig.\ \ref{fig:numodel}(b)) is compared to that of the real data (Fig.\ \ref{fig:numodel}(a)) with $\Delta t =1$[min] in Fig.\ \ref{fig:compfil}.
The fluctuation of the simulation data is consistent with that of the real data. 
In particular, we see good agreement around $\nu\simeq\epsilon$ as a result of the linear approximation of the interaction function.
Figure\ \ref{fig:numodel}(c) shows $\nu(t)$ of the simulation without  the interaction, i.e. $a=0$. 
The quantity $\nu$ fluctuates freely, which is inconsistent with the real data.

\subsection{Analytical approach}

We can solve the time-evolution equation\ (\ref{eq:nuevo}) analytically in some cases.
Let us define
\begin{equation}
\omega(t) \equiv \nu(t)-\epsilon
\end{equation} 
and
\begin{equation}
\left\{
\begin{array}{llllll}
\omega &=& \omega(t),&\omega'&=&\omega(t+\Delta t),
\\
F &=& F(t),&F'&=&F(t+\Delta t),
\\
c&=&c(\Delta t).&&&
\end{array}
\right.
\end{equation}
Equation\ (\ref{eq:nuevo}) is then reduced to 
\begin{equation}
\omega'=c \omega+F
\end{equation}
Assume that the probability of $\omega$ having a value in $\omega\sim\omega+{\rm d}\omega$ is $P_\omega (\omega)$.
The joint probability of $\omega'$ having a value in $\omega'\sim\omega'+{\rm d}\omega'$ and $F'$ having a value in $F'\sim F'+{\rm d}F'$ is given by
\begin{eqnarray}
 P_{\omega',F'}(\omega',F'){\rm d}\omega'{\rm d}F'&=& P_\omega (\omega) P_F (F'){\rm d}\omega{\rm d}F'  
\nonumber 
\\
&=&
\frac{1}{c}P_\omega(\frac{\omega'-F'}{c})
P_F(F'){\rm d}\omega'{\rm d}F',
\end{eqnarray}
where $P_F(F)$ is the probability of $F$ having a value in $F\sim F+{\rm d}F$.
The probability density function of $\omega'$ is thus given by
\begin{equation}
P_{\omega'}(\omega')=\frac{1}{c}\int P_\omega(\frac{\omega'-F'}{c})P_F(F'){\rm d}F'.
\end{equation}
The characteristic function of $\omega'$ is the Fourier transform
\begin{equation}
\tilde{P}_{\omega'}(k) 
=
\int P_{\omega'}(\omega') e^{i \omega'k} {\rm d}\omega'
=
\tilde{P}_{\omega}(ck) \tilde{P}_{F}(k),
\end{equation}
where $\tilde{P}_{\omega}$ and $\tilde{P}_{F}$ are the Fourier transforms of $P_{\omega}$ and $P_F$, respectively.
Then we obtain
\begin{eqnarray}
\tilde{P}_{\omega(t)}(k)
&=&
\tilde{P}_{\omega(t-\Delta t)}(ck)\tilde{P}_{F(t - \Delta t)}(k) 
\nonumber \\ 
&=&
\tilde{P}_{\omega(0)}(c^N k)\prod_{n=0}^{N}\tilde{P}_{F}(c^n k) 
\nonumber \\ 
&=&
\tilde{P}_{\omega(0)}(c^N k)\prod_{i=1}^{3}\prod_{n=0}^{N}\tilde{P}_{f_i} (c^n k) 
\nonumber \\ 
&=& 
\prod_{i=1}^{3}\prod_{n=0}^{N}\tilde{P}_{f_i}(c^n k),
\end{eqnarray}
where $t=N\Delta t$. We here assumed $P_{\omega(0)}(\omega)=\delta(\omega)$ and hence $\tilde{P}_{\omega(0)}(k)=1$.

The above argument shows the essential reason of the sharp peak and fat tails in Fig.\ \ref{fig:fils}. 
If we had $c=1$, or $a=0$ (without the interaction), the noise $F$ at every time step would accumulate in $\omega$ and the probability density function of $\omega=\nu-\epsilon$ would be Gaussian due to the central limit theorem. If we have $c<1$, or $a>0$ (with the interaction), the noise at the past time steps decay as $c^n$.
The largest contribution to $\omega$ comes from the noise one time step before, which is a fat-tail noise \cite{econophysics}.

As a special case, if the noises $f_i(t)$ obey a L$\acute{\rm e}$vy distribution of the same index $\alpha$ and the same scale factor $\gamma$, namely if
\begin{equation}
\tilde{P}_{f_i}(k)=e^{-\gamma|k|^\alpha }\quad\mbox{for all}\ i,
\end{equation}
the distribution of $\omega$ is also a L$\acute{\rm e}$vy distribution of the same index $\alpha$ and a different scale factor $\gamma'$ given by
\begin{equation}
\gamma'=\frac{3}{1-\{c(\Delta t)\}^\alpha}\gamma. 
\end{equation}

\section{Summary}

We first showed that triangular arbitrage opportunities exist in the foreign exchange market. 
The rate product $\mu$ fluctuates around a value $m$.
Next, we introduced a model including the interaction caused by the triangular arbitrage transaction. 
We showed that the interaction is the reason of the sharp peak and the fat-tail property of the distribution of the logarithm rate product $\nu$.
Finally we showed that our model is solvable analytically in some cases.
\section*{Acknowledgments}
One of the authors (Y. A.) is grateful to Mr. Tetsuro Murai and Mr. Takashi Anjo for their interesting comments.

\begin{figure}[p]
(a) 
  \begin{center}
    \includegraphics[width=8cm]{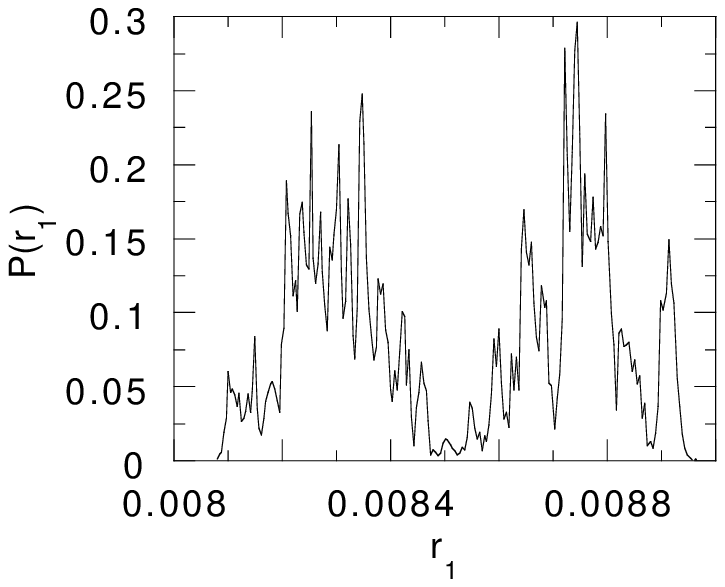} \\
  \end{center}

(b)
 \begin{center}
   \includegraphics[width=8cm]{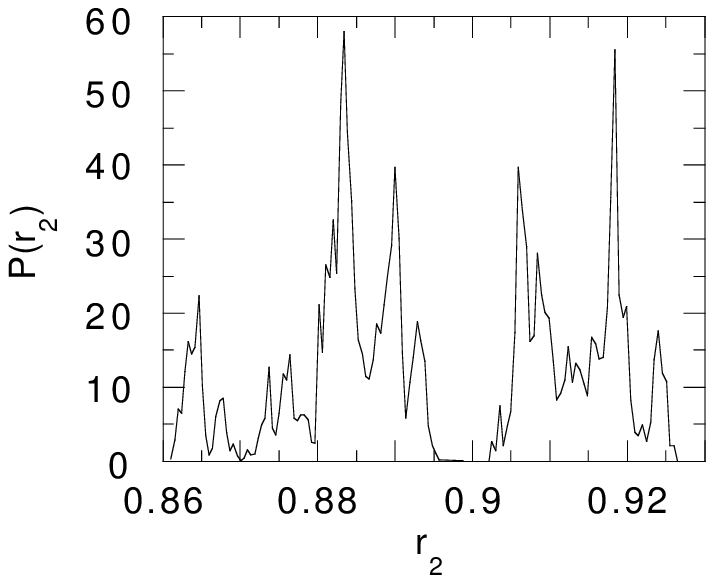} 
 \end{center}
\end{figure}

\begin{figure}[p]
(a)\\
 \begin{center}
  \includegraphics{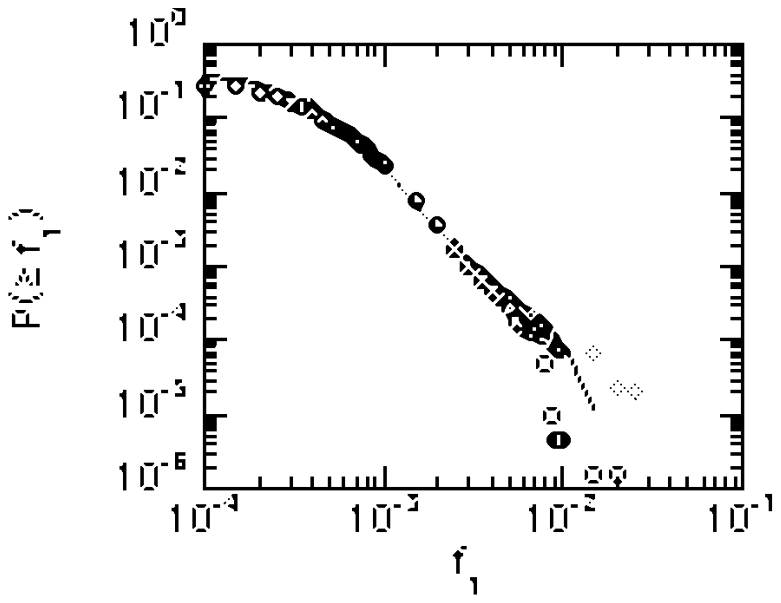} 
 \end{center}

(b)
  \begin{center}
   \includegraphics{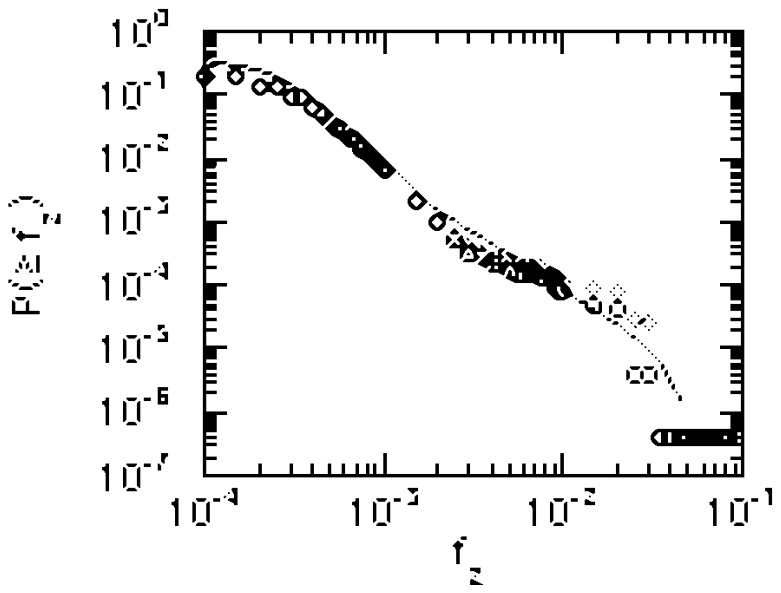} 
 \end{center}
\end{figure}

\begin{thebibliography}{1}

\bibitem{moosa}
I. Moosa, \lq Triangular Arbitrage in the Spot and Forward Foreign Exchange Markets,\rq {\it Quantitative Finance {\rm {\bf 1} 387-390 (2001)}}. 

\bibitem{mandelbrot}
B. B. Mandelbrot, \lq The Variation of Certain Speculative Prices,\rq {\it J. Business {\rm {\bf 36}, \rm 394-419 (1963)}}. 

\bibitem{econophysics} 
R. N. Mantegna, H. E. Stanley, {\it An Introduction to Econophysics: Correlations and Complexity in Finance} \rm(Cambridge University Press, Cambridge, 1999) pp.64-67.

\bibitem{arch}
R. F. Engle, \lq Autoregressive Conditional Heteroskedasticity with Estimates of the Variance of U.K. Inflation,\rq \it Econometrica \rm\bf 50\rm , 987-1002 (1982).

\bibitem{garch}
T. Bollerslev, \lq Generalized Autoregressive Conditional
  Heteroskedasticity,\rq \it J. Econometrics \bf 31\rm, 307-327 (1986).

\bibitem{bouchaud} 
J. P. Bouchaud, M. Potters, {\it Theory of Financial Risks: From Statistical Physics to Risk Management} \rm(Cambridge University Press, Cambridge, 2000) p.34 and pp.56-63.

\bibitem{takayasu}
H. Takayasu, M. Takayasu, M. P. Okazaki, K. Marumo and T. Shimizu in {\it Fractal Properties in Economics,} Paradigms of Complexity, World Scientific, ed. Miroslav M. Novak, (2000) pp.243-258

\bibitem{mantegna}
R. N. Mantegna, \lq Fast, Accurate Algorithm for Numerical Simulation of L$\rm\acute{e}$vy Stable Stochastic Processes,\rq \it Phys. Rev. \rm E\bf\ 49\rm , 4677-4683 (1994).

\bibitem{voit} 
J. Voit, {\it The Statistical Mechanics of Financial Markets} \rm(Springer, Berlin, 2001) pp.101-103.

\end{thebibliography}
\end{document}